\documentclass[]{aa} 

\def\mean#1{\left< #1 \right>}
\usepackage{subcaption}
\usepackage{natbib}
\usepackage{graphicx}
\usepackage{txfonts}

\begin{document} 

\title{A long-term study of AGN X-ray variability }
\subtitle{Structure function analysis on a ROSAT-XMM quasar sample}
                  \author{R. Middei
          \inst{1,  \thanks{riccardo.middei@uniroma3.it}}
          \and
          F. Vagnetti
          \inst{2}
          \and
          S. Bianchi
          \inst{1}
          \and
          F. La Franca
          \inst{1}
          \and
          M. Paolillo
          \inst{3,4,5}
        \and
         F. Ursini 
           \inst{1,6} 
}

   \institute{Dipartimento di Matematica e Fisica, Universit\`a degli Studi Roma Tre, via della Vasca Navale 84, 00146 Roma, Italy
             \and
             Dipartimento di Fisica, Universit\`a di Roma "Tor Vergata", via della Ricerca Scientifica 1, 00133 Roma, Italy
             \and
             Dip.di Fisica Ettore Pancini, Universit\`a di Napoli Federico II, C.U.
Monte Sant'Angelo, Via Cinthia, 80126 Napoli, Italy
            \and
         INFN Sezione di Napoli, Via Cinthia, 80126 Napoli, Italy
          \and
Agenzia Spaziale Italiana – Science Data Center, Via del Politecnico snc,
00133 Roma, Italy
           \and
            Univ. Grenoble Alpes, IPAG, F-38000 Grenoble, France
             }


 \abstract{}{}{}{}{} 
 
  \abstract
   {Variability in the X-rays is a key ingredient in understanding and unveiling active galactic nuclei (AGN) properties. In this band, flux variations occur on short timescales (hours) as well as on larger timescales. While short timescale variability is often investigated in single source studies, only a few works are able to explore flux variation on very long timescales.}
   {This work aims to provide a statistical analysis of the AGN long term X-ray variability. We study variability on the largest time interval ever investigated for the 0.2-2 keV band,  up to approximately $ 20$ years rest-frame for a sample of 220 sources. Moreover, we study variability for 2,700 quasars up to approximatley eight years rest-frame in the same (soft) band.}
   {We built our source sample using the 3XMM serendipitous source catalogue data release 5, and data from ROSAT  All Sky Survey Bright and Faint source catalogues. To ensure that we selected AGN only, we used the Sloan Digital Sky Survey quasar catalogues data releases 7 and 12. 
  Combining ROSAT and XMM-Newton observations, we investigated variability using the structure function analysis which describes the amount of variability as a function of the lag between the observations.}
   {Our work shows an increase of the structure function up to 20 years. We find no evidence of a plateau in the structure function on these long timescales.}
   {The increase of the structure function at long time lags suggests that variability in the soft X-rays can be influenced by flux variations originated in the accretion disk or that they take place in a region large enough to justify variation on such long timescales.}

   \keywords{galaxies:active – quasars:general – X-rays:galaxies
     }

   \maketitle

\section{Introduction}

Active galactic nuclei (AGN) are extremely bright sources shining in the local as well as in the far Universe in several bands of the electromagnetic spectrum. Variability in AGN is observed both in the emission lines and continuum emission, thus it is a defining characteristic for these sources \citep[e.g.][]{Matt63a, Elli74}. Flux variability occurs on different timescales from hours \citep{ponti12}, up to several years \citep[e.g.][]{de-V03a, Vagn11a}.
Variability studies can be useful in discovering fundamental properties of the central engine of AGN. Indeed,  many authors have used flux variations to constrain the size of the emitting region, the emission mechanisms, as well as the process originating the variability itself. Studying single objects flux variations on tenths of ks \cite[e.g.][]{ponti12} it is found that X-ray variability is already present at such short timescales.

Variability observations on short timescales suggest that X-ray emission takes place in a very small region \citep[e.g.][]{Mush93} close to the central black hole (BH), the so called corona whose geometry and size are still matter of debate, \citep[e.g.][]{Haar91, Haar94}.
Single-source analyses are often performed using the power spectral density analysis (PSD). However, this analysis needs well sampled light curves that are only available for few objects. The PSD approach finds a break at high frequencies that is connected with the BH mass \cite[e.g.][]{Papa04,Onei05,Uttl05,McHa06}, while, even if a low frequency break is expected, we have experimental evidences of it only for a couple of sources: for example, Ark 564  \cite[][]{Poun01, Papa02, McHa07} and NGC 3783 \citep{Mark03}.\\
\indent As reported by several authors \citep[e.g.][]{,Zhan11,Youn12,Shem14}, AGN display flux variations also on larger temporal scales.
It is possible to investigate variability on larger time intervals using the structure function (SF) analysis \citep[e.g.][]{Simo85, Vagn11a}. SF works on the time domain and can be used in statistical studies even when objects have poorly sampled light curves, see details in Sect. 3. In this context, a statistical approach to characterise variability \citep{ Vagn16} is useful to point out some averaged properties shared by AGN. Further to this, an ensemble study minimises the effect of anomalous sources if they are present. To date, we know that X-ray variability increases as a function of the time interval at which we investigate the sources. This means, in an average sense, that if we compare flux measures performed in two different epochs we can expect that, the larger is the delay $\tau$ between the two observations, the larger will be the difference between the two fluxes.

\indent Variable X-ray absorption, which is commonly observed in Seyfert galaxies, can also contribute to the X-ray variations. In particular, changes of the X-ray absorbing column density are ubiquitous in type 2, with timescales ranging from hours to years \citep{Risa02}. Moreover, X-ray absorption variability has also been found in a number of type 1 \cite[see e.g.][]{Mark14, Pucc07, Turn08}.
The shorter-term variations in covering fraction of partially covering absorbers are generally ascribed to eclipses by clouds from the broad-line region \citep[e.g.][]{Risa11}. Longer-term variations both in type 1 and type 2, on timescales up to a few years, could be produced by a pc-scale torus with a clumpy structure,  with type 2 having a much higher probability \cite[e.g][]{Mark14}.  We do not investigate the absorbtion variability contribution in the present work, however a possible signature of this may be searched for by selecting a subsample of AGN inclined at large viewing angles, adopting angular indicators, as, for example, proposed by \cite{Shen14}.

\indent The source sample we analyze in this work is based on serendipitous observations performed by XMM-Newton and ROSAT data. ROSAT  observations are needed in order to investigate variability on long timescales. In fact, other authors have already compared ROSAT and XMM-Newton data for different purposes \citep[e.g.][]{Saxt11, Stro16}; comparing data from these two satellites, we are now able to perform a long-term variability analysis for a sample of 2,856 AGN observed from early nineties up to 2015.\\
\indent This paper is organised as follows: in Sect. 2 we describe the data extracted from the archival catalogues and we compare data by ROSAT and XMM-Newton. Section 3 contains the analysis we perform and our results. In Sect. 4 we discuss important timescales for variability and we report our conclusions. In Sect. 5 we summarise this work. Throughout the paper we adopt the standard cosmological $\Lambda CDM$ framework identified by $\Omega_{\Lambda}=0.7$, $\Omega_{M}=0.3$, and $H_0=70$~Km/Mpc/s.

\section{Data}

In this work we have used X-ray data obtained by the XMM-Newton and ROSAT satellites.
The 3XMMSSC-DR5\footnote{http://xmmssc.irap.omp.eu/} \citep{Rose16} contains 565,962 X-ray detections belonging to 396,910 single sources. The observations were performed between February 2000 and December 2013, thus data cover a large time interval, see Fig. 1. Furthermore, this catalogue is very suitable for ensemble variability studies since it contains 70,453 sources observed at least twice (up to 48 times) for a total of 239,505 multi-epoch observations.
\begin{figure}[htbp]
        \centering
        \includegraphics [width=3.5in]{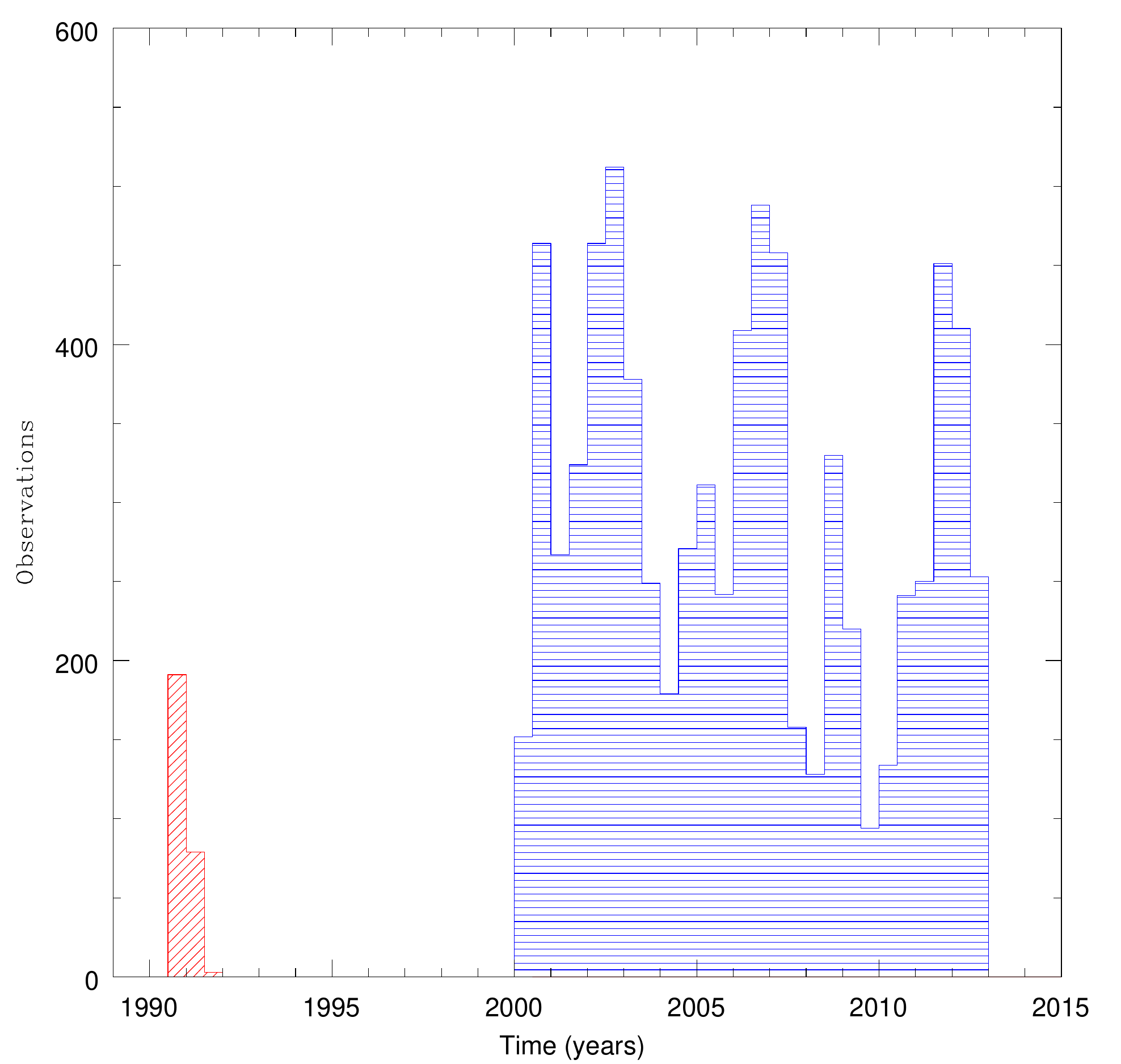}
        \caption{Time distribution of the observations coming from the XMM-Newton mission (blue, horizontal shading) and ROSAT (red, slanted shading).}
        \label{fig:1}   
\end{figure}

As mentioned above, XMM-Newton observations were carried out over a period of 13 years, however because of the redshift, the maximum length for the rest-frame light curves of these sources is approximately eight years, as is shown in Fig. 2. To enlarge our investigating temporal window we have taken advantage of the RASS-BSC \citep{Voge99} and RASS-FSC \citep{Voge00}, which store respectively 18,806 and 105,924 X-ray detections. ROSAT observations were obtained between July 1990 and August 1991, thus comparing this information with data performed by XMM-Newton we can study variability up to time lags of about $20$ years.
 
The 3XMMSSC-DR5 as well as the RASS catalogues contain X-ray detections with no information concerning their nature. Therefore to extract from the sample only AGN we cross-correlated these data with the quasar catalogues by the Sloan Digital Sky Survey (SDSS), the SDSS-DR7Q \citep{Schn10a} and the SDSS-DR12Q \citep{Pari16}. The SDSS quasar catalogues do not include classes of active galactic nuclei such as BL Lacertae and Type 2 objects, thus they allow us to build a homogeneous sample of type 1 AGN.
To cross-match catalogues we took advantage of the software TOPCAT\footnote{http://www.star.bris.ac.uk/$\sim$mbt/topcat/}       \citep[][]{Topcat1, Topcat2}.

 We obtained our quasar sample extracting  from the 3XMMSSC-DR5 only observations of AGN. To do this we cross-matched the serendipitous source catalogue by XMM-Newton with both SDSS-DR7Q and SDSS-DR12Q using a 5 arcsec cross-correlating radius.
Moreover, we accounted for the quality of the observations selecting only the observations with a $SUM\_FLAG$ value smaller than three, as suggested by the 3XMMSSC team.
This procedure led us to obtain 14,648 matches corresponding to 2700 multiply observed AGN (7,837  observations) and 6,801 pointed only once. The sample of sources with multiple observations corresponds to the Multi-Epoch XMM Serendipitous AGN Sample, MEXSAS\footnote{Available at:\\ http://vizier.cfa.harvard.edu/viz-bin/VizieR?-source=J/A+A/593/A55}, already presented in \citet{Vagn16}. In the following we will refer to the extended sample containing both multiple and single observations as eMEXSAS.

To test our matches we used a larger radius of 10 arcsec for which the cross-correlation algorithm produces 15,095 matches indicating a possible incompleteness of the order of 3$\%$.
Furthermore, we checked our cross-correlations, giving a set of spurious coordinates shifted by 1 arcmin in declination to the eMEXSAS entries and we obtain 44 matches indicating that our sample contains false occurrencies of the order of 0.3$\%$.

\begin{figure}[htbp]
        \centering
        \includegraphics [width=3.5in]{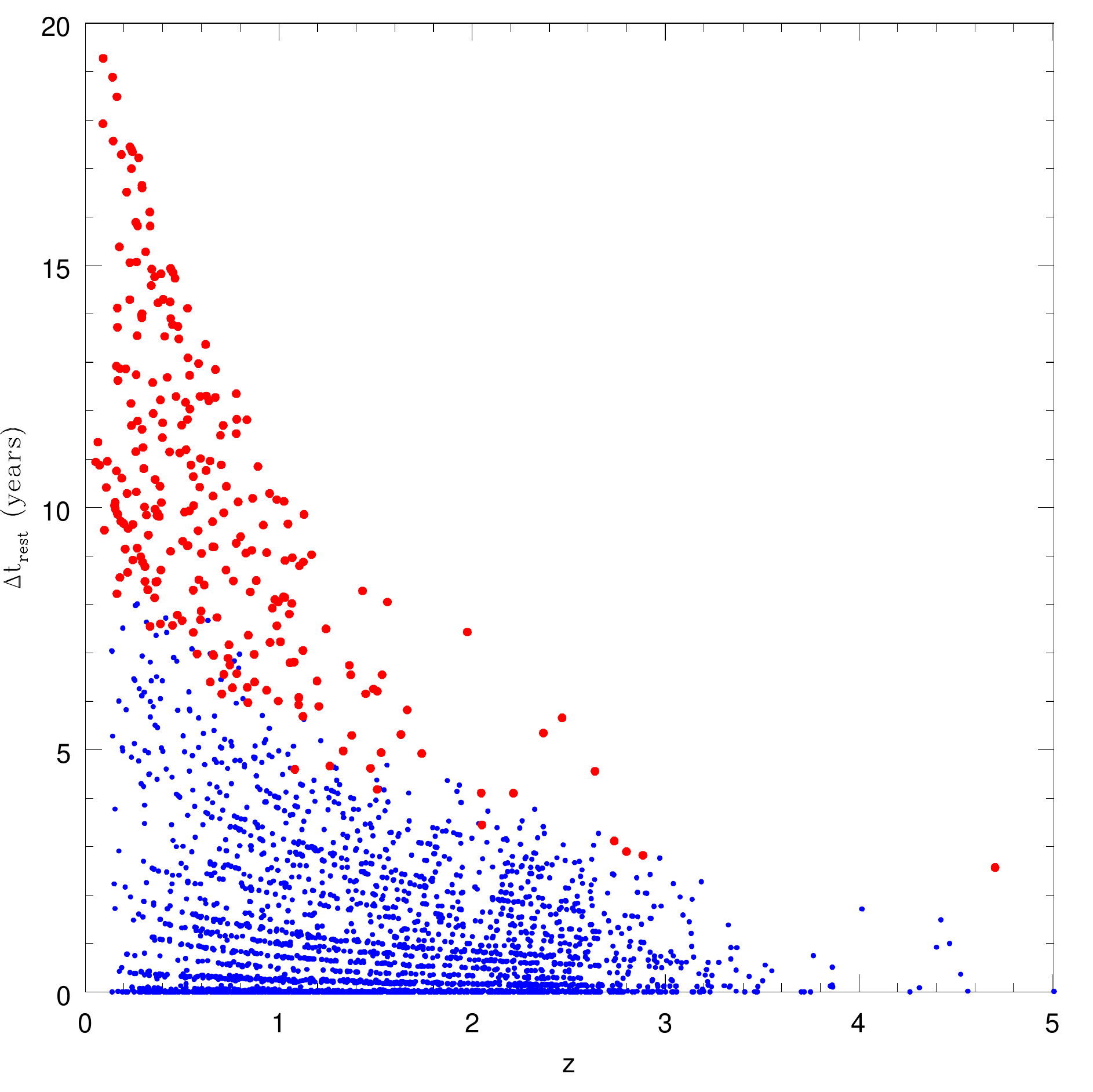}
        \caption{Length of the light curve in the rest frame, $\Delta t_{rest}$, for each source, as a function of the redshift. Small blue dots: XMM-Newton data alone, larger red points: combined XMM-Newton+ROSAT data.}
        \label{fig:1}   
\end{figure}

Then, we cross-correlated our extended sample eMEXSAS with the RASS-BSC and the RASS-FSC adopting a radius of 30 arcsec. This leads us to obtain 281 sources observed by the two satellites for a total of 490 available observations. This cross-match is a crucial point in our work, then we repeat the cross-correlations using a set of false coordinates, shifting by one degree in declination finding six matches. This suggests that our sample contains a $\sim$1$\%$ of possible contamination within the adopted matching radius.
Then we checked our cross-correlations again using the column of quasar catalogues SDSS-DR12 and SDSS-DR7 in which a ROSAT count rate coming from the SDSS identification is available. We discarded a few sources from our sample for which there is no SDSS identification. This reduced our sample to 273 single sources observed both by XMM-Newton and ROSAT. For these sources the maximum length of the light curves $\Delta t_{rest}$ is as long as 19 years, see Fig. 2.

Data used in this work are provided by two missions with different characteristics \citep[e.g.][respectively for ROSAT  and XMM-Newton]{true82, stru01}, therefore it is not possible to compare their measurements in a straightforward way. In the following we describe the steps we performed to produce a reliable comparison of our data. The first issue we face is due to the different energy bands used by ROSAT and XMM-Newton to scan the sky. The ROSAT  X-ray telescope, in fact, operated in the 0.1-2.4  keV band while XMM-Newton performs observations in the energy band 0.2-12 keV. Due to the smaller energy range of ROSAT we can study variability at long time lags only in the soft X-ray band. 
In order to make the measure performed compatible in these different observational bands, we made use of the online tool WebPIMMS\footnote{https://heasarc.gsfc.nasa.gov/cgi-bin/Tools/w3pimmw/w3pimms.pl}. 

To convert ROSAT count rate to fluxes we set WebPIMMs as follows: Galactic absorption $N_H=3\times10^{20}~\textrm{cm}^{-2}$, source model power law with a $\Gamma=1.7$ \citep[the same values are used in the 3XMMSSC-DR5 by][]{Rose16}, input energy range 0.1-2.4 keV and output energy range 0.2-2 keV. 
These flux measures have to be compared with those obtained from 3XMMSSC-DR5. The serendipitous catalogue reports flux measures in the small energy bands $EP1\_FLUX$ (0.2-0.5 keV), $EP2\_FLUX$ (0.5-1.0 keV), $EP3\_FLUX$ (1.0-2.0 keV). We combined the flux measures in the energy bands EP1, EP2, EP3 adopting the procedures described by \cite{wats09} in their Appendix 4, D.4, getting, for all the sources in our sample, a soft integrated flux in the energy band 0.2-2 keV.

\section{Analysis}

Structure function is a powerful method used in several electro-magnetic bands by many authors, in the Optical-UV  \cite[e.g.][]{Trev94a, Kawa98a, de-V03a, Baue09a} as well as in the radio domain \citep[e.g.][]{Hugh92a}. Since 2011 SF analysis has also been used in the X-ray band where authors produced variability studies for statistically rich AGN populations \citep{Vagn11a,Vagn16}. Structure function works in the time domain and it is suitable for ensemble studies. This is a real advantage because, even if for single source variability characterization SF as well as  PSD need richly sampled light curves, for an ensemble approach SF can be adopted also when only poorly sampled light curves are available. 
Structure function describes variability giving a measure of the mean change between two observations separated by a time lag $\tau$.
Different mathematical formulations for SF have been used in the past \citep[e.g.][]{Simo85,di-C96a}, however in this work we adopt the following definition for this variability estimator:
\begin{equation}
SF(\tau)=\sqrt{\mean{[\log f_X (t+\tau)-\log f_X(t)]^2}-\sigma^2_{noise}}~.
\end{equation}
In Eq.1 $\tau$ is the lag elapsing between the two available flux measures and $\sigma_{noise}^2$ is the the quadratic contribution of the photometric noise to the observed variations: 
\begin{equation}
\sigma_{noise}^2=\mean{\sigma^2_{n}(t)+\sigma^2_{n}(t+\tau)}~,
\end{equation}
where $\sigma_n$ is the photometric error of the logarithmic flux at a given epoch. The average quantities in Eq. 1 and Eq. 2 are computed within an appropriate bin of time lag around $\tau$.

Our sample of sources spreads on a large redshift interval, thus we compute the time lag in the source rest-frame,

\medskip
\begin{equation}
\tau_{rest}=\frac{\tau_{obs}}{1+z}~.
\end{equation}
\medskip
\begin{figure}[htbp]
        \centering
        \includegraphics [width=3.5in]{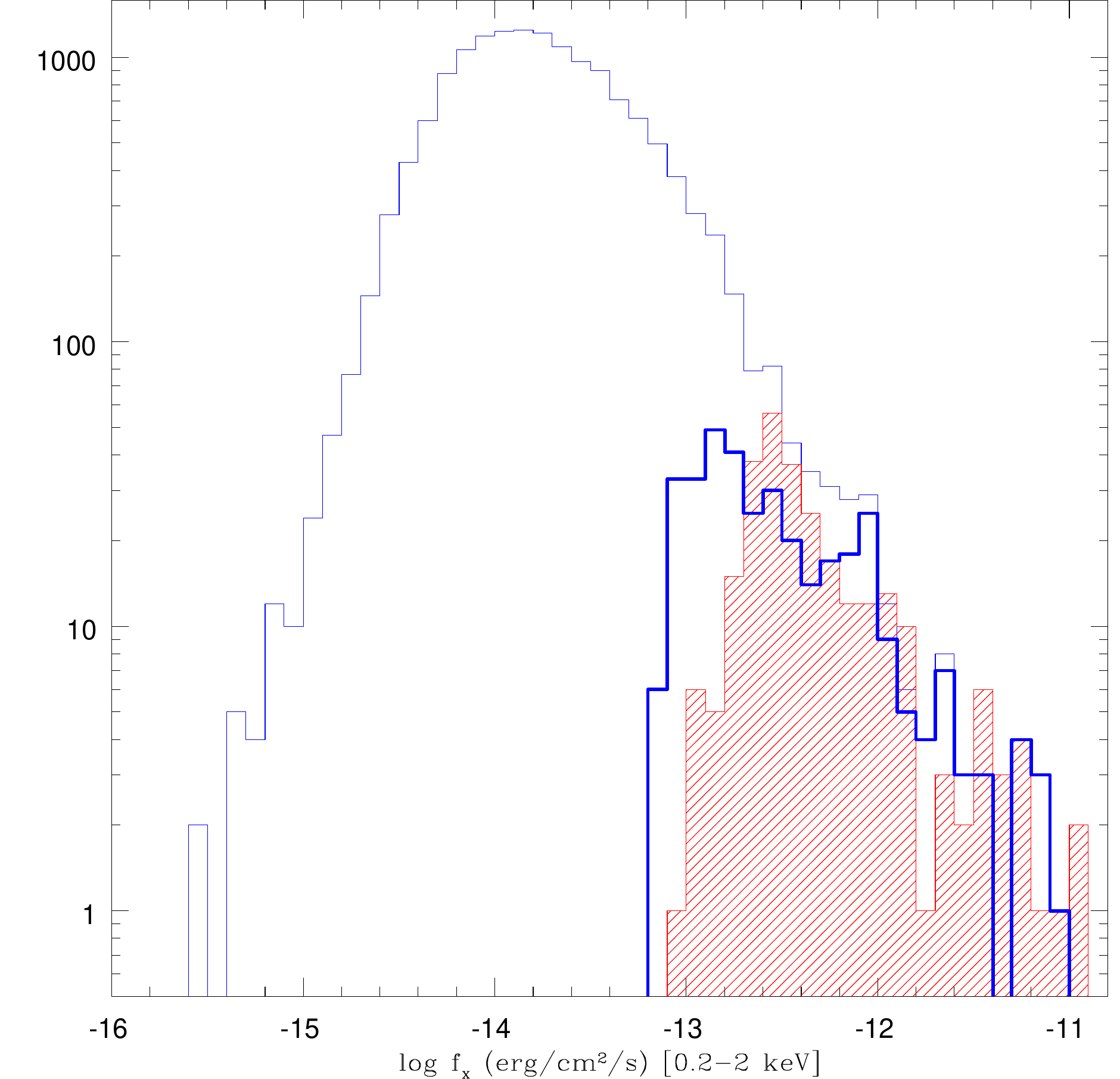}
        \caption{Flux distribution in the 0.2-2 keV band for the observations belonging to the different subsamples. Thin blue histogram: MEXSAS; thick blue histogram: XMM-Newton observations brighter than $7.5\times10^{-14}$~erg/cm$^2$/s for the sources observed by ROSAT; red shaded histogram: ROSAT fluxes.}
        \label{fig:1}   
\end{figure}

Our soft long-term SF analysis takes advantage of eMEXSAS and the AGN selection observed by ROSAT.
SF is computed using all the light curves for the sources belonging to eMEXSAS and extending these light curves for the combined XMM-Newton and ROSAT relevant subsample.  

Observations performed by a single mission allow us to compute structure function straightforwardly but in the case of combined light curves, data are obtained from two satellites with different instrumental properties. In particular for this study we take care of the different sensitivities of the two space observatories. For quasars observed by the two missions, ROSAT fluxes have values spanning between $6.6\times10^{-14}$ erg/cm$^2$/s and $1.2\times10^{-11}$ erg/cm$^2$/s. On the other hand, due to its more recent technology, fluxes detected by XMM-Newton are also fainter, down to $2.5\times10^{-16}$ erg/cm$^2$/s. Structure function, as shown by Eq. 1, describes variability comparing flux measures performed in different epochs, thus the different sensitivities of the two satellites could introduce a bias, which would give an overestimate of the large flux variations (in this case the increase of the SF might be an artifact), being on average ROSAT detections much brighter than XMM-Newton measurements. Moreover, a possible bias for the cross calibrations between XMM-Newton and ROSAT observations has been discussed by \cite{Shem14} who estimate that this does not exceed the $10\%$ of the flux measures. Following the procedure outlined by \citet{Saez12}, we added in quadrature to the errors of each measured flux a 10\% error. This increases slightly the estimate of the flux errors for the brightest sources, however its effect on our variability analysis is very small (see below).

\begin{figure}[h!]
        \centering
        \includegraphics [width=3.5in]{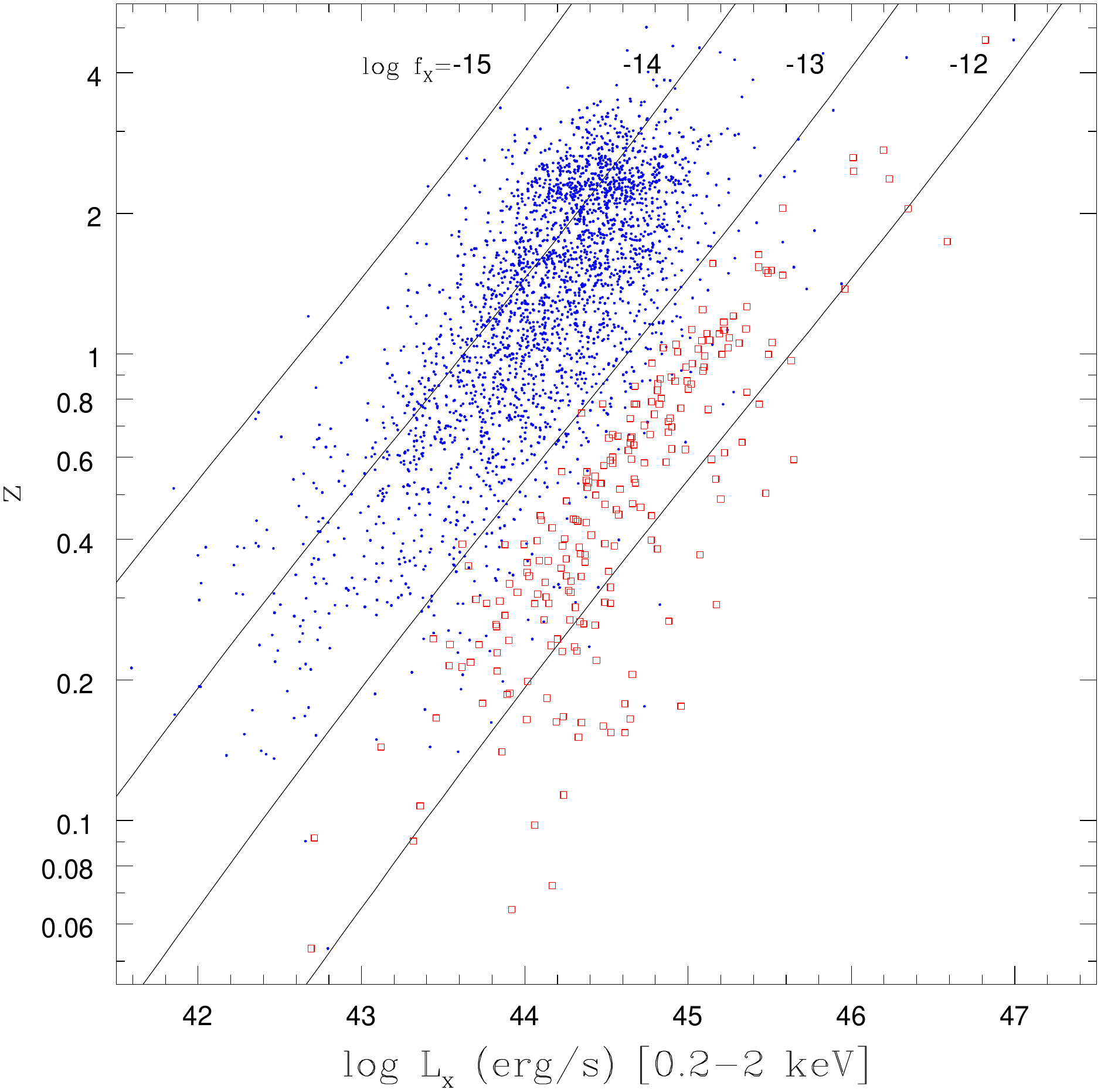}
        \caption{Distribution of the sources for which SF is computed in the $L_X$-z plane. The small blue dots represent the average values of the X-ray luminosity for sources with multiple XMM-Newton observations (MEXSAS), while empty squares represent the luminosity and redshift computed from ROSAT observations.
}
        \label{fig:1}   
\end{figure}

Furthermore, ROSAT  measures are systematically sampling earlier times. This means that when we evaluate SF, fluxes at earlier epochs are in average larger with respect to later epoch measures performed by XMM-Newton. Therefore light curves tend to have a decreasing behaviour, in average, for these sources.

This leads us to define a common threshold for the XMM-Newton and ROSAT fluxes such that the corresponding flux distributions span a similar interval. The adopted threshold ($7.5\times10^{-14}$~erg/cm$^2$/s) optimises the overlap between the two distributions, see Fig. 3 where the flux distribution of the overall MEXSAS sample is also shown for comparison. This discards 53 ROSAT sources associated with XMM-Newton fluxes fainter than the threshold, so that our subsample of combined light curves is reduced to 220 sources. In Fig. 4 we show on the $L_X-z$ plane the distribution of the sources we study in this work constituted by MEXSAS (2700 sources, small blue dots) plus the 220 from ROSAT (empty red squares).
We note that the two samples span a similar range in luminosity, thus the luminosity dependence of X-ray variability \citep[see e.g.][]{Vagn16} should not introduce a difference between the two samples.

\begin{figure}[ht!]
        \centering
        \includegraphics [width=3.5in]{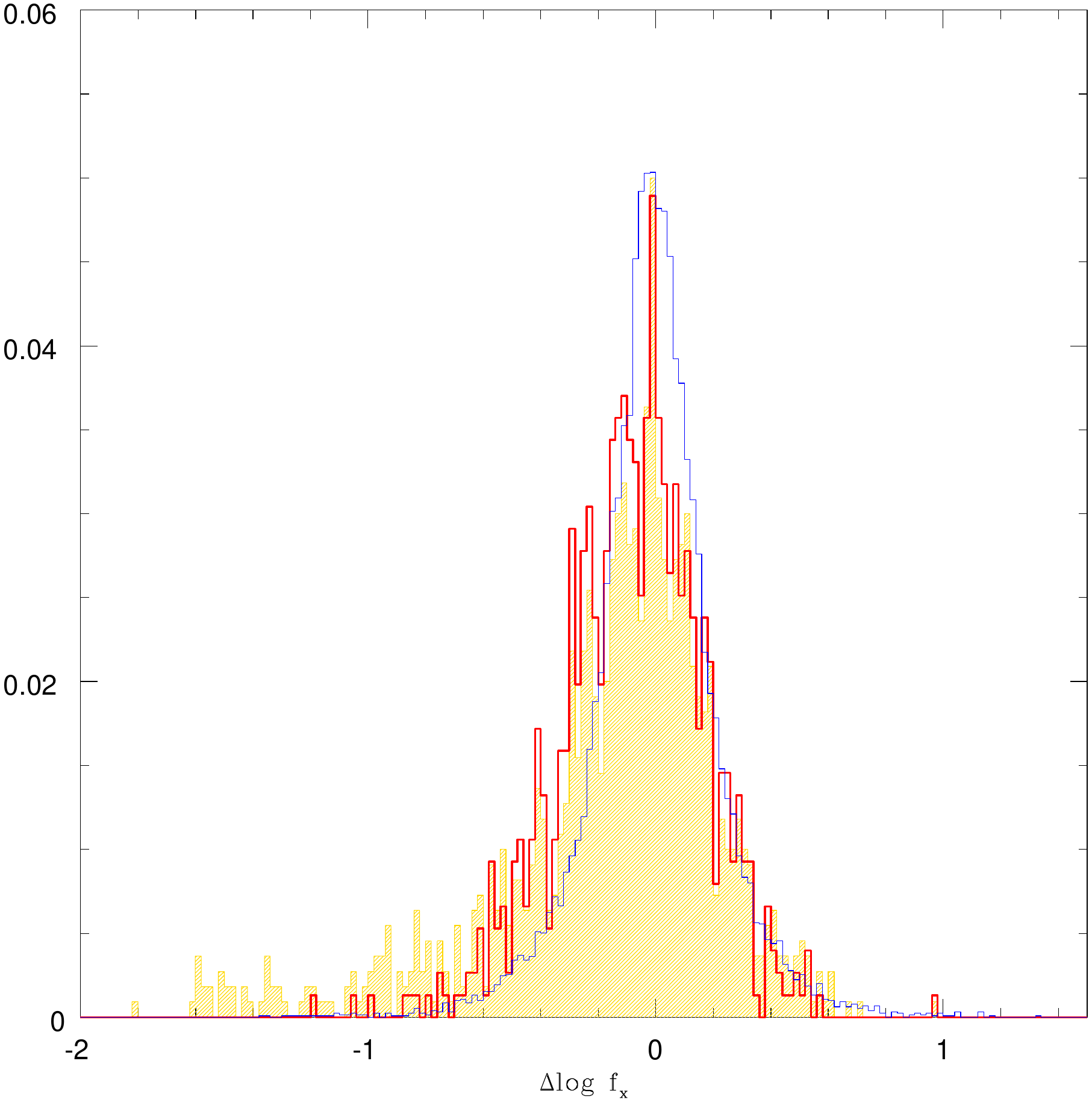}
        \caption{Distribution of the logarithmic flux variations, normalised to the total number. Yellow shaded histogram: combined ROSAT  XMM-Newton light curves including all the data. Red histogram: the same as before but removing the XMM-Newton fluxes fainter than $7.5 \times 10^{-14}$~erg/cm$^2$/s. Blue histogram: XMM-Newton-only light curves for the MEXSAS sample.}
        \label{fig:1}   
\end{figure}

We now discuss the effect of the different flux limits of the ROSAT and XMM-Newton catalogues displaying in Fig. 5 the distributions of the logarithmic flux variations expected for our combined lights curves in two cases: if we include XMM-Newton fluxes fainter than the adopted threshold we find an asymmetric distribution with an excess tail towards negative large variations (yellow shaded in Fig. 5); cutting instead the XMM-Newton measures to fluxes larger than the threshold, the asymmetry of the resulting distribution is strongly reduced (red histogram in Fig. 5). For comparison we display in the same figure also the corresponding distribution for the MEXSAS sample alone.

We show in Fig. 6 the distribution of the rest frame time lags for sources belonging to MEXSAS (blue histogram) and to the ROSAT and XMM-Newton combined sample (red shaded histogram). As Fig. 6 displays, the contribution of the ROSAT data is particularly important for the larger rest-frame timescale ($\tau \gtrsim$ 8 years) while, for smaller time lags, this contribution is almost negligible compared to data of MEXSAS.\\
\begin{figure}[htbp]
        \centering
        \includegraphics [width=3.5in]{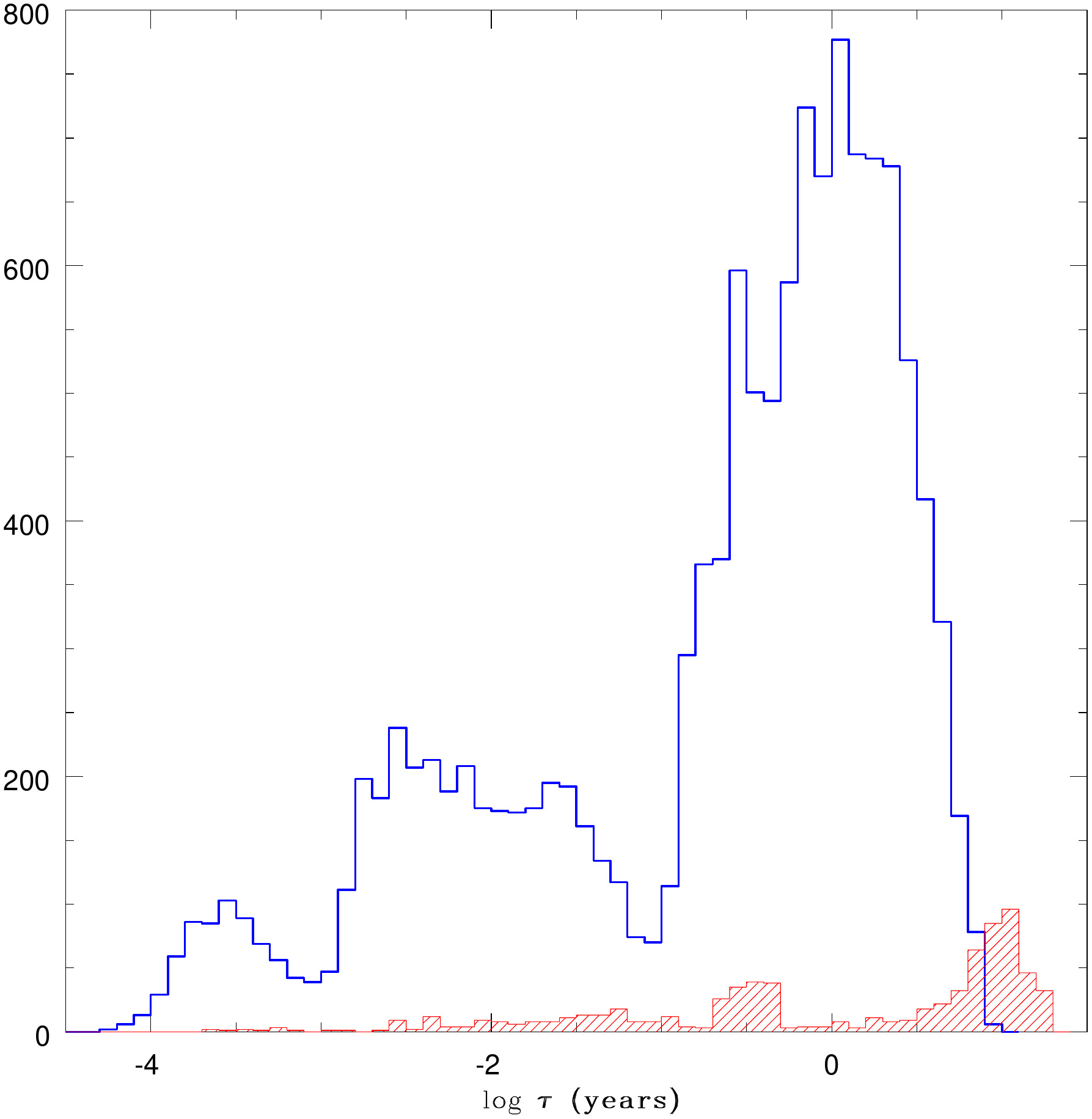}
        \caption{Distribution of the rest frame time lags for MEXSAS sources (in blue) and for the combined XMM-Newton (with fluxes brighter than $7.5 \times 10^{-14}$~erg/cm$^2$/s.) and ROSAT data in red. Combined observations extend the distribution to larger time lags.}
        \label{fig:1}   
\end{figure}

\indent Finally, using Eq. 1 we are able to compute the longest structure function ever performed for ensemble X-ray variability studies.
\begin{figure}[htbp]
        \centering
        \includegraphics [width=3.5in]{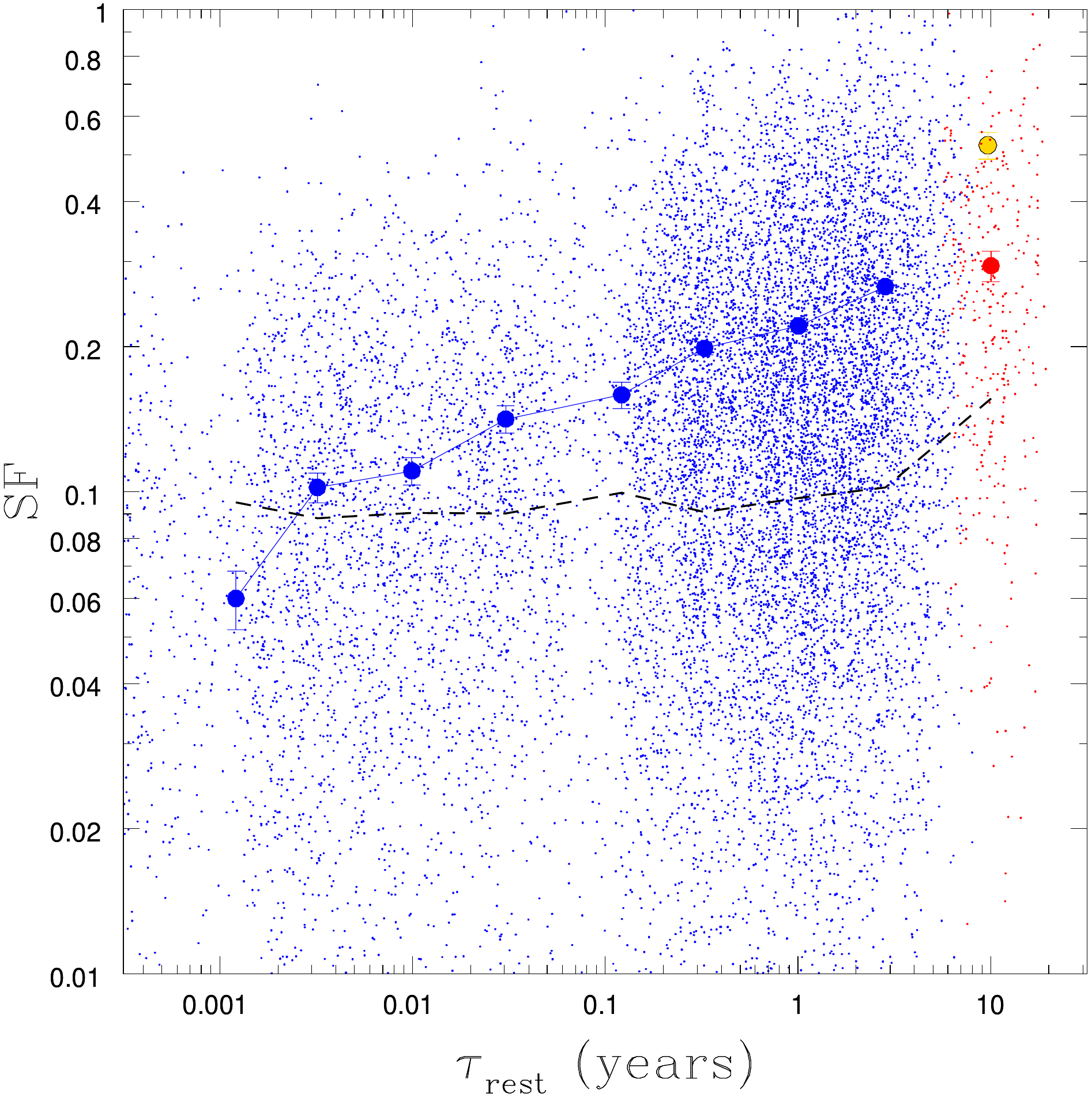}
        \caption{Ensemble structure function for the sources studied in this work. Blue points represent the averaged SF values and take advantage only of XMM-Newton observations, while the red point is obtained also using data from ROSAT. The red point refers to 220 AGN pointed by both the satellites and is computed cutting the XMM-Newton fluxes to the ROSAT flux threshold as discussed in the text. The yellow, black circled point is computed without removing the XMM-Newton fluxes fainter than the threshold, and it strongly overestimates the amount of variability. The black dashed line is the contribution of the noise (as defined in Eq. 2) for the whole investigated time-lag interval. Small dots represent the variations for the individual pairs of measurements contributing to the SF and the colours emphasise the different subsamples used during the SF calculation.}
        \label{fig:1}   
\end{figure}
In Fig. 7 we present the binned structure function showing the contributions of the combined ROSAT-XMM-Newton data in red and only in the bin $\tau_{rest}\gtrsim8$ years while, for the bins at smaller time lags, we plot only the dominant contribution coming from MEXSAS. For comparison, we also draw an estimate of the SF in the last bin using the full XMM-Newton light curves without removing the fluxes fainter than the previously discussed threshold.
However, this corresponds to an overestimate ($\sim 75\%$) of the variability, while a more conservative estimate is also indicated on the figure, computed with only the ROSAT and XMM-Newton measurements brighter than the threshold. Figure 7 also shows the contribution of the measurements errors (dashed line); the previously discussed intercalibration errors are included in our analysis, slightly increasing the error contribution (by $\sim 5\%$) for the combined XMM-Newton/Rosat bin. However the effect on the SF (according to Eq. 1), is negligible, $\sim 1\%$.

The computed SF still increases for time lags as long as twenty years and we do not find evidence for a flattening.
The increase of the structrure function can be fitted by a power-law $SF\propto\tau^b$, with $b=0.154\pm0.011$ for the eight bins spanned by XMM-Newton data, and $b=0.153\pm0.010$ including the ROSAT bin in its conservative estimate discussed above.
 Our results are in agreement with those of  \citet{Zhan11}, who reported a lack of a low-frequency break in the PSD of local AGNs, using RXTE light curves on timescales of around ten years. This nicely complements our results on a large sample of high-$z$ objects, suggesting a similar long-term variability trend in local AGNs and quasars.

\section{Discussion}
The SF presented in the previous section does not show a flattening at long timescales which, in principle, could be expected considering the finite size of the emitting region.  As discussed by, for example, \cite{Czer06}, X-ray variability on short timescales can be originated by intrinsic fluctuations of the X-ray source while, at longer timescales, the observed X-ray variations can be affected by optical-UV fluctuations.
Indeed, also optical SF computed by \cite{de-V05} shows a similar behaviour with an increase at long time lags. However, \citet{MacL12} report a SF flattening at about two years, although the authors do not rule out a continuing but slower rise.

 It is possible to interpret our result considering previous works by \cite{Lyub97} and \cite{Arev06}. These authors discuss about the propagating fluctuations model in which optical long term variations originated in the external region of the disk propagate through it affecting variability in the X-ray band. Flux variations acting on short timescales produced in the disk regions closer to the central engine are therefore modulated by long term variations.
We note that the influence of disk-born fluctuations on the X-ray variations is also supported by the variability of the X-ray-to-optical spectral index $\alpha_{ox}$, which increases at year-long timescales \citep{Vagn10,Vagn13}.
In this scenario, the size of the disk could leave some track on our SF, therefore we want to estimate the characteristic timescales in the external part of the accretion disk. 

Following \citet{Czer06}, we focus on the dynamical timescale, which is shorter than the viscous and thermal timescales, and can be expressed for a Keplerian optically thick and geometrically thin disk as follows:
\begin{equation}
\tau_{dyn}=7\times10^{-5}({R}/{R_g})^{3/2}{M}/{M_\sun} ~{\rm yrs},
\end{equation}
where $R_g=GM/c^2$ is the gravitational radius and $M$ is the black hole mass.
The external size of the accretion disk is not well known, however it can be estimated following the work by \cite{Coll01}, who argue that for large radii ($\gtrsim 1000 ~R_g$) the disk becomes self gravitating. They estimate a critical radius $R_{crit}$ at which the disk becomes gravitationally unstable, fragmenting and possibly giving rise to the broad line region.
$R_{crit}$ depends on the Eddington ratio and on BH mass. In the interval $0.01~<L/L_{bol}<1$ the dependence on the mass can be approximated as
\begin{equation}
R_{crit}/R_g=3\times 10^7 ({M}/{M_\sun})^{-0.46}~,
\end{equation}
which corresponds to a disk size $R=1.5\times10^{-6} (M/M_\odot)^{0.54}$ pc.
At this point we can combine the equations to get a self-consistent estimate for the dynamical time at the outer border of the accretion disk $R_{crit}$ as a function of the BH mass:
\begin{equation}
\tau_{dyn}=0.35 ({M}/{M_\sun})^{0.31} ~{\rm yrs}.
\end{equation}

Equation 6 suggests timescales of several decades for black hole masses of $10^6-10^9$ $M_\odot$. Therefore an increase of SF at twenty years appears compatible with the estimated disk sizes.

The geometrical and physical properties of the X-ray-emitting region are not yet fully understood. On the one hand, the power law-like continuum emission in hard X-rays is generally believed to originate in a hot and compact corona, likely located in the inner part of the accretion flow \citep[e.g.][]{Reis13}. However, on the other hand, an excess of emission below 1-2 keV above the extrapolated high-energy power law is commonly observed in the spectra of AGN \citep[e.g][]{Walt93, Bian09}. Albeit the nature of this so-called soft X-ray excess is uncertain \citep[e.g.][]{Done12}, a possible explanation is thermal Comptonization by a warm, optically thick medium \citep[e.g.][]{Magd98, Petr13}.
In particular, \cite{Petr13} studied the high-energy spectrum of  Mrk 509 in detail, using the data from a long, multiwavelength campaign \citep{Kaas11}. This source showed  a correlation between the optical-UV and soft (< 0.5 keV) X-ray flux, but no correlation between the optical-UV and hard (> 3 keV) X-ray flux \citep{Mehd11}.  Indeed, \cite{Petr13} found that the spectrum is well described by a two-corona model: a warm ($kT$$\sim$1 keV), optically thick ($\tau\sim15$) corona responsible for both the optical-UV emission and the soft X-ray excess, and  a hot ($kT$$\sim100$ keV), optically thin ($\tau\sim0.5$) corona responsible for the hard X-ray emission.
Moreover, the authors affirm that the warm corona agrees with a slab geometry, thus it may cover a large fraction of the disk  \citep{Petr13}.
Other authors \citep[e.g.][]{Jani01, Roza15} suggest that this warm corona could be the upper layer of the disks itself. According to this view, the presence of an extended warm corona could further increase the timescales for variability in the X-ray band.

\section{Summary}
In this work we have investigated the long-term AGN variability for the soft X-ray band using SF analysis. In order to do this we extracted AGN sources (observed at least twice) from data coming from XMM-Newton and ROSAT  observations. Data from these two satellites cover a time interval of about $\sim22$ years (Fig. 1), corresponding to $\sim20$ years interval in the rest frame (see Fig. 2). Before performing our analysis we check our sample and we correct offsets due to the different satellite properties and sensibilities. After cleaning the data, we computed structure function in Fig. 7 that refers to a sample of 2864 sources in total and covers a time interval of $\sim20$ years rest frame. Our analysis does not show any evidence of a plateau.
Our SF shows an increase at long time lags which can be understood in a scenario in which relevant timescales are due to the variations occurring in the external region of the accretion disk or a large emitting region originates the soft photons.

\begin{acknowledgements}

We thank Francesco Tombesi, Massimo Cappi, Mauro Dadina \& Andrea Marinucci, for useful discussions. 
We thank the referee for his/her useful comments which helped to improve the quality of this work.
We acknowledge funding from PRIN/MIUR-2010 award
2010NHBSBE. This research has made useof data obtained from the 3XMM XMM-Newton serendipitous source catalogue
compiled by the 10 institutes of the XMM-Newton Survey Science Centre selected by ESA. Funding for SDSS-III has been provided by the Alfred P. Sloan
Foundation, the Participating Institutions, the National Science Foundation, and
the U.S. Department of Energy Office of Science. The SDSS-III web site is
http://www.sdss3.org/. SDSS-III is managed by the Astrophysical Research Con-
sortium for the Participating Institutions of the SDSS-III Collaboration including
the University of Arizona, the Brazilian Participation Group, Brookhaven Na-
tional Laboratory, Carnegie Mellon University, University of Florida, the French
Participation Group, the German Participation Group, Harvard University, the
Instituto de Astrofisica de Canarias, the Michigan State/Notre Dame/JINA Par-
ticipation Group, Johns Hopkins University, Lawrence Berkeley National Labo-
ratory, Max Planck Institute for Astrophysics, Max Planck Institute for Extrater-
restrial Physics, New Mexico State University, New York University, Ohio State
University, Pennsylvania State University, University of Portsmouth, Princeton
University, the Spanish Participation Group, University of Tokyo, University of
Utah, Vanderbilt University, University of Virginia, University of Washington,
and Yale University.
\end{acknowledgements}

\thispagestyle{empty}
\bibliographystyle{aa}
\bibliography{middei16.bib}

\end{document}